\documentclass[useAMS,usenatbib]{mn2e}

\usepackage{graphicx}

\newcommand{\comment}[1]{}

\title[LUNASKA UHE neutrino flux limits for Cen A and the GC]{LUNASKA Experiment observational limits on UHE neutrinos from Centaurus A and the Galactic Centre}
\author[C.~W.~James et al.]{C.~W.~James$^{1,2}$\thanks{E-mail: C.James@astro.ru.nl},
R.~J.~Protheroe$^{2}$,
R.~D.~Ekers$^{3}$,
J.~Alvarez-Mu\~niz$^{4}$, \newauthor 
R.~A.~McFadden$^{3,5}$,
C.~J.~Phillips$^{3}$, 
P.~Roberts$^{3}$,
J.~D.~Bray$^{2,3}$\\
$^{1}$IMAPP, Radboud University, Nijmegen, The Netherlands\\
$^{2}$School of Chemistry \& Physics,  Univ.\ of Adelaide, Australia.\\  
$^3$Australia Telescope National Facility, CSIRO Astronomy \& Space Science, Epping, Australia.\\
$^4$Dept.\ Fisica de Particulas \& IGFAE, Univ. Santiago de Compostela, Spain. \\
$^5$School of Physics, Univ.\ of Melbourne, Australia.}

\begin{document}

\date{}

\pagerange{\pageref{firstpage}--\pageref{lastpage}} \pubyear{2010}

\maketitle

\label{firstpage}

\begin{abstract}
We present the first observational limits to the ultra-high
energy (UHE) neutrino flux from the Galactic Center, and from
Centaurus A which is the nearest active galactic nucleus (AGN).  These
results are based on our ``Lunar UHE Neutrino Astrophysics using
the Square Kilometer Array'' (LUNASKA) project experiments at the
Australia Telescope Compact Array (ATCA).  We also derive limits
for the previous experiments and compare these limits with
expectations for acceleration and super-heavy dark matter models of the origin of UHE cosmic rays.
\end{abstract}


\maketitle


\section{Introduction}

Arrival directions of the UHE cosmic rays (CR) detected by the
Pierre Auger experiment above $5.6\times 10^{19}$~eV have been
found to be statistically correlated with positions of nearby AGN
\citep{AugerScience07}, and a few of the arrival directions
appear to be clustered around Centaurus A, our nearest active
galactic nucleus at a distance of $\sim$3.7~Mpc.  This has led to
speculation that Centaurus A may be responsible for some of the
UHE CR.  However, the flux is extremely low, and so the nature of
the sources of UHE CR remains at present unknown.

As well as observing UHE CR directly, an alternative means of
exploring the origin of UHE CR is to search for UHE neutrinos.
Cosmic rays of sufficient energy will interact (e.g.\ via pion
photo-production) with photons of the 2.725~K cosmic microwave
background (CMB) radiation, with the resulting energy-loss
producing a cut-off in the spectrum at around $\sim10^{20}$~eV
from a distant source (\citealt{Greisen}; \citealt{Zatsepin_Kuzmin}). These same
interactions produce ``cosmogenic'' neutrinos from the decay of
unstable secondaries
(\citealt{Stecker1973Ap&SS..20...47S}; \citealt{Stecker1979ApJ...228..919S}; \citealt{BerezinskyGZK}; \citealt{ProtheroeJohnson96}; \citealt{Engel01}). As well as these
cosmogenic neutrinos, UHE neutrinos are also expected to be
produced by acceleration and super-heavy dark matter (SHDM) sources of UHE CR, and some
information on the CR spectrum at the sources is imprinted on the
spectrum of cosmogenic neutrinos \citep{Protheroe04}. Of course,
neutrinos are not deflected by magnetic fields, and so should
point back to where they were produced. See
\citet{ProtheroeClay2004} and \citet{Falcke2004_SKAscienceCase}  for
reviews of UHE CR production scenarios and radio
techniques for high-energy cosmic ray and neutrino astrophysics.

\subsection{The Lunar Cherenkov Technique}

In our present work we use the lunar Cherenkov technique
\citep{Dagkesamanskii}, in which the Moon is used as a UHE
neutrino target and Earth-based radio telescopes are used to
detect coherent radio Cherenkov emission produced by
neutrino-induced cascades in the lunar regolith.  A high-energy
particle interacting in a dense medium will produce a cascade of
secondary particles which develops an excess negative charge by
entrainment of electrons from the surrounding material and
positron annihilation in flight.  The charge excess is 
proportional to the number of particles in the electromagnetic
cascade, which in turn is proportional to the energy
of the primary particle.  \citet{Askaryan62} (see also \citealt{Askaryan65}) first noted
this effect and predicted the  Cherenkov emission process in dense
dielectric media to be coherent at radio frequencies where the
wavelength is comparable to or larger than the dimensions of the shower, and
this effect has been confirmed experimentally
\citep{Saltzberg_GorhamSAND01}. At wavelengths comparable to the
width of the shower, the coherent emission is in a narrow cone
about the Cherenkov angle, while for wavelengths comparable to the
shower length the coherent emission is nearly isotropic \citep{Alvarez-Muniz06}.

The lunar Cherenkov technique aims to utilize the outer layers of
the Moon, nominally the regolith which is a sandy layer of ejecta
covering the Moon to a depth of $\sim$10~m, as a suitable medium
to observe the Askaryan effect. Since the regolith is
comparatively transparent at radio frequencies, coherent
Cherenkov emission from cascades due to sufficiently high-energy neutrino
interactions in the regolith should be detectable as
nanosecond-scale pulses by Earth-based radio-telescopes.

\begin{figure}
\begin{center}
\includegraphics[width=0.49\textwidth]{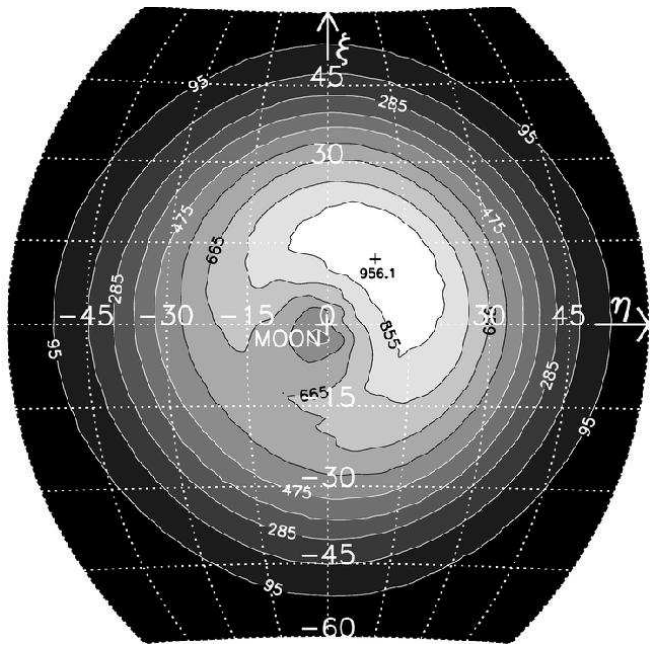}
\caption{Contours of effective area (km$^2$) as a function of UHE neutrino
arrival direction for $10^{23}$~eV neutrinos for limb-pointing configuration (May
2008). The `+' marks the position of peak effective area; the
Moon is at the center $(\eta,\xi)$=(0,0); telescope pointing direction $(\eta,\xi)$=(0.183$^\circ$,0.183$^\circ$).}
\label{ATCA_instant}
\end{center}
\end{figure}

\section{Description of the Experiment}

The aim of the LUNASKA project is to develop further the lunar
Cherenkov technique for UHE neutrino astronomy, and to influence
the design of the Square Kilometer Array
(\verb!http://www.skatelescope.org/!) so that UHE neutrino
observations may be possible.  Our experiment was carried out at
the ATCA which is an aperture synthesis telescope located at
latitude $-$30$^\circ$ near Narrabri, Australia.  It consists of
six identical $22$~m dishes of which we have used three with
baselines ranging from $\sim$100~m to $\sim$750~m.  The ATCA has
a half-power beam-width that matches the lunar disk at 1.4~GHz,
and it provided us with $600$~MHz (1.2--1.8 GHz) of bandwidth.
Being an array the ATCA also provided both strong timing
discrimination against terrestrial radio-frequency interference
(RFI), and gave a large effective area and high sensitivity while
seeing the entire moon \citep{Ekers08}.  In order to perform a
search for nanosecond-duration lunar Cherenkov pulses, we had to
build specialized hardware, including analogue de-dispersion
filters as such pulses suffer dispersion in the Earth's
ionosphere -- our experiment is the first to coherently correct
for this before detection in real-time.  To detect and store
candidate events in real time we used field-programmable gate
array (FPGA) based analog-to-digital converters developed by the
Australia Telescope National Facility, each of which could
digitize and perform simple logic on two data streams at a rate
of $2.048$~GHz.  The signal was passed to both a running buffer
of length 256 samples and a real-time trigger algorithm. We
triggered independently, with a maximum rate of 1040 Hz, at each
antenna. On fulfilling the trigger conditions, the buffer was
returned to the control room and recorded.  We calibrated the
system sensitivity using the thermal emission from the lunar
disk, and the system clocks at each antenna using correlated
emission from the quasar 3C273.  Full details of the experiment
are given by \citet{James_PhD2009} and \citet{James_PRD2009}.

The observations described here cover two observing periods,
February and May 2008.  The February 2008 observations were
tailored to target a broad ($\ga 20^{\circ}$) region of the sky
near the Galactic Center, both a potential accelerator of UHE CR
and also a potential source of UHE CR, gamma-rays and neutrinos
through its dark matter halo.  Based on simulation results
(\citealt{JamesProtheroe09a}; \citealt{JamesProtheroe09b}), we
pointed the antennas at the lunar center in February to achieve
the greatest total effective aperture and sensitivity to an
isotropic or very broadly-distributed flux. Our May 2008
observing period targeted Centaurus A only, and by pointing the
telescopes at the portion of the lunar limb closest to Centaurus
A we achieved maximum sensitivity for this source.  We recorded a
total of 98307 3-fold coincidences within 4-$\mu$s windows, the
majority of which, based on statistical arguments, could not come
from random noise.  Fitting these to far-field sources, $\sim 60$
appeared to come from the direction of the Moon.  However, our
0.5 nanosecond timing resolution allowed us to show that these
were all either near-field in origin, or were identifiable under
visual inspection as long-duration, narrow-band RFI signals with
little or no time-structure, occurring during short periods of
intense RFI.  Thus we excluded all candidates as being of lunar
cosmic-ray or neutrino origin;
cosmic rays interacting with lunar regolith with favourable
geometry could also produce observable lunar Cherenkov emission.

\begin{figure*}
\begin{center}
\includegraphics[width=0.7\textwidth]{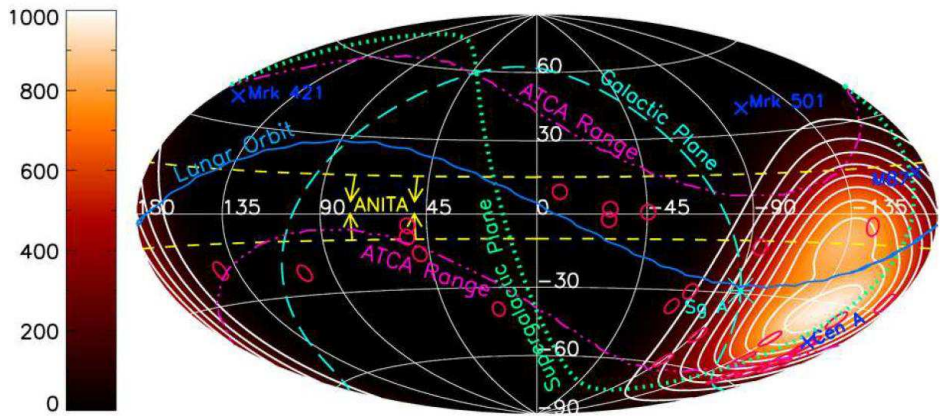}
\caption{The exposure (km$^2$-days) of the 2008 LUNASKA UHE $\nu$
detection experiment using the ATCA to $10^{23}$~eV
neutrinos. The small circles show the directions of UHE CR events
above $5.6\times 10^{19}$~eV detected by Auger
\citep{AugerScience07}.  The positions of the Galactic Centre and
four important AGN are indicated by crosses labelled
``Sgr~A$^*$'', ``Mrk 421'', ``Mrk 501'', ``M87'' and ``Cen~A''.
The Galactic plane and Supergalactic plane are indicated by
long-dashed and dotted curves, respectively.  The region of sky
accessible to lunar Cherenkov observations using the ATCA is
between the two chain curves labelled ``ATCA RANGE''.  In the
declination range $-$10$^{\circ} \la \delta \la $+15$^{\circ}$
(between the short dashed curves labelled ``ANITA'') the exposure
of ANITA far exceeds that of other experiments, but for clarity
has not been plotted.}
\label{ATCA_combined_exposure}
\end{center}
\end{figure*}

We calculated the effective areas as described by
\citet{JamesProtheroe09b}. This is shown in Fig.\
\ref{ATCA_instant} for $10^{23}$~eV neutrinos and the
limb-pointing configuration, and we see that the sensitivity has
a characteristic `kidney' shape peaking at $\sim$$15^\circ$ away
from the Moon along the line extending from the Moon's center to
the telescope pointing position on the lunar limb.  For the
center-pointing configuration (not shown), the sensitivity
pattern forms an annulus, with peak exposure around
$15^{\circ}$--$20^{\circ}$ degrees from the Moon.  For both
configurations, the sensitivity pattern broadens with increasing
primary particle energy as the increased strength of the pulses
produced allows the telescopes to be sensitive to a wider range
of interaction geometries.  Combining the instantaneous aperture,
e.g.\ as shown in Fig.\ \ref{ATCA_instant}, with the known
telescope-pointing positions on the Moon and the Moon's position
itself at the time of observing, allows us to a calculate the
exposure $[A \times t](E_\nu;\alpha,\delta)$ (effective area-time
product) as a function of celestial coordinates
$(\alpha,\delta)$.  The exposure from our LUNASKA ATCA
observations to UHE neutrinos is shown in Fig.\
\ref{ATCA_combined_exposure} for $10^{23}$~eV.  The nominal declination range of
the ANITA observations is also given, and in this range the
exposure of ANITA (not shown) dominates.  The concentration of
exposure about Centaurus A, and the broad Galactic Centre region
(nominally Sagittarius A), both of which are outside ANITA's
sensitive declination range, is due to the targeting of these
regions in our experiment by a careful choice of observing times.

\section{Results}

The experiment with the greatest exposure to UHE neutrinos in the
energy range above $10^{21}$~eV applicable here is the Antarctic
Impulsive Transient Antenna Experiment (ANITA) (\citealt{ANITA};
\citealt{ANITA_2010}) but this experiment is only sensitive to
the declination range $-$10$^{\circ} \la \delta \la
$+15$^{\circ}$ (indicated in Fig.~\ref{ATCA_combined_exposure}).
The IceCube experiment is primarily sensitive to the northern
hemisphere, but has nevertheless been able to set limits up to
$10^{18}$~eV for Centaurus A and the Galactic Center
\citep{IceCube_CenA}.  The Pierre Auger observatory, although
designed as a cosmic ray detector, is also sensitive to
neutrinos, and a preliminary limit has been published for
neutrinos from Centaurus A with energies up to $10^{20}$~eV
\citep{Auger_CenA_nu}.  
Other experiments for which we can obtain
a neutrino flux limit from Centaurus A above $10^{17}$~eV are the
pioneering experiment at Parkes \citep{Parkes}, the Goldstone
Lunar UHE Neutrino Experiment (GLUE) \citep{GLUE}, the NuMoon
observations using the Westerbork interferometer \citep{NuMoon},
and the Radio Ice Cherenkov Experiment (RICE) \citep{RICE}. The
directional dependence of the exposures of the original Parkes
experiment and GLUE were calculated previously by
\citet{JamesProtheroe09b}.

The NuMoon lunar Cherenkov observations at Westerbork are
reported by \citet{NuMoon}. While neither Centaurus A nor the
Galactic Centre was explicitly targeted,  lunar Cherenkov
experiments operating at an order of magnitude lower frequency
than ATCA, such as the Westerbork array, are expected to be more
sensitive at energies above $10^{23}$ eV \citep{NuMoon}.  We
simulated the sensitivity to isotropic $10^{23}$ eV neutrinos for
the NuMoon experimental configuration of two fan beams on
different sides of the Moon, each with 4 frequency sub-bands of
width 65 MHz centred on 123, 137, 151 and 165 MHz covering $1/3$
of the Moon.  Adopting a 10 m thick regolith with a denser
sub-regolith layer, and using the triggering criteria as
described by \citet{NuMoon} we found a sensitivity a factor of
$\sim 10$ lower than that calculated by \citet{NuMoon}.  
We are able to closely reproduce the published NuMoon result by
instead simulating one continuous bandwidth with a $240$~kJy
threshold (artificially improving the sensitivity by a factor of
$5$) and letting the depth of the top regolith layer become
infinite (gaining a further factor of $2$).

Using the method of \citet{JamesProtheroe09b}, but taking account
of the fan beams of the Westerbork antennas, the frequency
sub-bands and triggering criteria, our simulations show that for
NuMoon the instantaneous neutrino directional aperture rises
rapidly from zero at $0^{\circ}$ from the Moon to a peak at $\sim
40^{\circ}$ away from the Moon, and then drops to zero at $\sim
100^{\circ}$ away from the Moon.  To calculate the limit for
Centaurus A we obtained the observation dates from
\citet{Buitink09}, and on two of these dates the Moon happened to
be $\sim 45^\circ$ from Centaurus A.  Because only the
observation dates (not the times) were available we assumed
uniform lunar coverage by the two beams, but reduced by $1/3$
since each beam covered $\sim 1/3$ of the Moon.  We calculated
the neutrino sky coverage of NuMoon by integrating the
instantaneous neutrino directional aperture for the positions of
the Moon on the observation dates and obtained the NuMoon
sensitivity for $10^{23}$ eV neutrinos from Centaurus A.  We
obtained the sensitivity for the Galactic Centre using the same
method.

RICE was a Cherenkov radio experiment embedded in Antarctic ice
at the South Pole and had a much lower neutrino energy threshold
than the three lunar Cherenkov experiments, albeit with only a
slowly-increasing exposure with neutrino energy. However, it had
a very long observation time of several years as compared to
several days for the lunar Cherenkov experiments and this
compensates for its lower instantaneous effective aperture.  RICE
was mostly sensitive to down-going neutrino events, and so to
declinations $\delta$$<$0$^\circ$.  The directional sensitivity
pattern at $10^{22}$~eV \citep{Besson08_private} shows that its
sensitivity to Centaurus~A and the Galactic Centre are
respectively 1.0 and 1.4 times its average sensitivity to a
source in the southern hemisphere.  Taking this directional
sensitivity as representative for all energies, and the energy
dependence of the exposure to a diffuse flux as given by
\citet{RICE} we can thus calculate the exposure of RICE to these
point sources. Since the isotropic exposure is only given to
$10^{22}$~eV in \citet{RICE}, in calculating the directional
exposure we use a log-linear scaling and assume an isotropic
exposure of $6.3 \times 10^{17}$~cm$^2$~s~sr to $10^{23}$~eV
neutrinos.

The individual and total exposures of the GLUE, RICE, NuMoon and
LUNASKA experiments to UHE $\nu$ from Centaurus A and Sgr A are
given in Table \ref{ATCA_exposures_table} for $10^{21}$,
$10^{22}$, and $10^{23}$~eV.  Model-independent 90\% confidence
limits to the neutrino flux at energy $E_\nu$ from putative point
sources at positions $(\alpha,\delta)$ can be obtained from the
exposure using $E_\nu F_\nu(E_\nu) \le 2.3/[A\times
t](E_\nu;\alpha,\delta)$.  The errors in the LUNASKA
exposures given in Table \ref{ATCA_exposures_table} reflect the
in-quadrature addition of two dominant factors - uncertainties in
the absolute detector thresholds, and the unknown properties of
the lunar regolith.  An estimate of the former is described in
our earlier publication \citep{James_PRD2009}. To estimate
uncertainties due to regolith properties, we use the measured
variation in moon rock radio absorption and density
\citep{OlhoeftStrangway75}.  The mare have a dense highly
radio-absorbent regolith, and we vary the fraction of the part of
the lunar surface relevant for lunar Cherenkov events due to
neutrinos from Centaurus A and the Galactic Centre.  For our
``best-case regolith'' estimate we use a $0$\% mare fraction,
since there are very few mare regions in the lunar South where
events from Centaurus A would be located.  For our ``worst-case
regolith'' estimate, we use a 30\% mare fraction, which is
approximately twice the lunar average and would correspond to the
near side of the Moon being uniformly covered with mare.  The
mare regions in our ``worst case regolith'' are modelled as
having density $\rho=3.0$~g/cm$^3$ and field attenuation length
$\ell=2.94$~m at $1$~GHz, while ``best case regolith'' is
modelled with $\rho = 1.8$ and $\ell = 22$~m.  Finally, we assume
a $(10 \pm 10)$\% increase in aperture due to the interactions of
secondary $\mu$ and $\tau$, as calculated for other similar
experiments by \citet{JamesProtheroe09a}.  Our error estimates do
not account for the extreme uncertainty in the UHE neutrino
interaction cross-sections -- these are so poorly understood at
these energies that determining them is an experimental goal as
much as a calculation uncertainty. Given the strong
proportionality between cross-section and effective
area/sensitivity for this and similar experiments
\citep{JamesProtheroe09a}, our flux limit is really a limit on
the flux---cross section product.

\begin{table*}
\renewcommand{\arraystretch}{1.6}
\caption{Experimental exposures (km$^2$-days) of GLUE, RICE, NuMoon (Westerbork) and the LUNASKA ATCA observations to UHE neutrinos at discrete energies from the Galactic Center and Centaurus A.  }
\begin{center}
\begin{tabular}{l  c c c r   c  c c c r }
\hline
$E_{\nu}$~~~~      & \multicolumn{4}{c}{Galactic Centre}             & ~   & \multicolumn{4}{c}{Centaurus A}   \\
(eV)		& GLUE	& RICE	& NuMoon & ATCA	 & ~  & GLUE	& RICE	& NuMoon & ATCA	 	 \\
\hline
$10^{21}$	& 0.5	& 47	& 0	& $3.2^{+2.7}_{-1}$	 & ~  	& 0.015	& 35	& 0	& $7.6^{+6.4}_{-2.5}$	 	\\
$10^{22}$	& 14	& 86	& 0	& $59^{+32}_{-15}$	 & ~  	& 2.1	& 65	& 0	& $122^{+66}_{-31}$		\\
$10^{23}$	& 175	& 157	& 73.2	& $450^{+102}_{-104}$	 & ~  	& 43	& 121	& 467	& $819^{+186}_{-189}$		\\
\hline
\end{tabular}
\end{center}
\label{ATCA_exposures_table}
\end{table*}

\begin{figure}
\begin{center}
\includegraphics[width=0.5\textwidth]{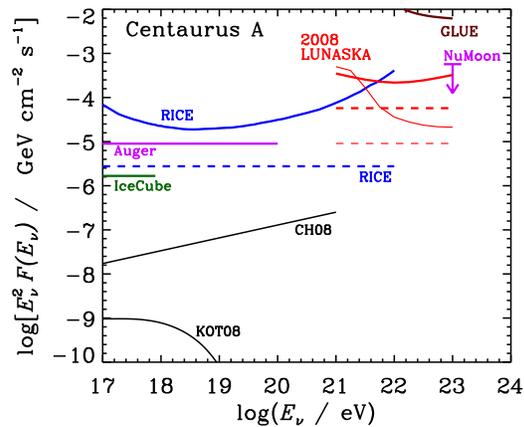}
\caption{Neutrino flux limits for Centaurus A from our 2008
LUNASKA ATCA experiments, and based on the GLUE, NuMoon and RICE, as
well as published results from IceCube and Auger experiments.
Limits are shown as integrated limits to a $E_{\nu}^{-2}$
spectrum (dashed lines), as well as model-independent limits
(solid curves) where available -- the NuMoon limit is
model-independent. For the LUNASKA ATCA limit we show results from
standard modeling (thick curve and thick dashed line) and using a
toy model of small-scale lunar surface roughness (thin curve and
thin dashed line) as described in Appendix B of
\citet{James_PRD2009} -- the abrupt transition near $2\times
10^{21}$ eV is a model artifact.  Neutrino flux predictions of
two AGN models for UHE CR production as labeled: KOT09
\citep{Kachelriess0805.2608}; CH08 \citep{CuocoHannestad023007}.}
\label{FluxLimitsCenA}
\end{center}
\end{figure}

\section{Summary and Conclusion}

In Fig.~\ref{FluxLimitsCenA} we show the all-flavor neutrino flux
limits for Centaurus A.  With Centaurus A only 3.7~Mpc away, and
with the pion photo-production energy-loss distance on the CMB
minimizing at $\sim$12~Mpc above $10^{11}$~GeV (e.g.\
\citealt{Protheroe04}) for rectilinear propagation one would
observe UHE CR almost unattenuated by pion photo-production
interactions on the CMB.
\citet{RiegerAharonian2009A&A...506L..41R} suggest that shear
acceleration along the kpc jet may accelerate protons beyond
$5\times 10^{19}$ eV.  The magnetic field of Centaurus A's giant
lobes \citep{Feain_CenA_LobesB_2009ApJ...707..114F} may also
provide an environment suitable for acceleration of UHE CR
(\citealt{BenfordProtheroe2008MNRAS.383..663B};
\citealt{Hardcastle_CenA_2009MNRAS.393.1041H}).

\citet{CuocoHannestad023007} have predicted the
flux of UHE neutrinos from the Centaurus A core
(`CH08' in Fig.~\ref{FluxLimitsCenA}) using a model of an
optically thick pion photo-production source described by
 \citet{MannheimProtheroeRachen01}. They assume that accelerated cosmic ray protons are
perfectly magnetically contained, and escape only through photo-hadronic interactions
which convert them to neutrons.  Under their model, the observed $E^{-2.7}$ spectrum of
cosmic rays requires a $E^{-1.7}$ proton injection spectrum within the source, which
would also produce a $E^{-1.7}$ spectrum of neutrinos; however, these assumptions may
break down towards the upper end of the energy range considered here.  They normalize
the CR flux by assuming that 2 of the Auger events above $5.6\times 10^{19}$ eV are from
Centaurus~A, and determine the relative normalization between the CR and neutrino fluxes
through Monte Carlo simulations of $p$--$\gamma$ interactions in the acceleration region
using the SOPHIA event generator \citep{SOPHIA2000CoPhC.124..290M}.

\citet{Kachelriess0805.2608} also assume a CR flux composed of protons, normalized with
the assumption that 2 of the UHE Auger events are from Centaurus~A.  They consider
several possible proton injection spectra, with acceleration occurring either in
regular electromagnetic fields close to the core of the AGN or through shock acceleration
in the jets, and predict the resulting neutrino and gamma-ray spectra for each.  The
jet acceleration scenarios are excluded by TeV gamma-ray data \citep{Kachelriess0904.0590},
and we plot (`KOT09' in Fig.~\ref{FluxLimitsCenA}) their result for acceleration near the
core with an $E^{-2}$ proton injection spectrum.

\begin{figure}
\begin{center}
\includegraphics[width=0.5\textwidth]{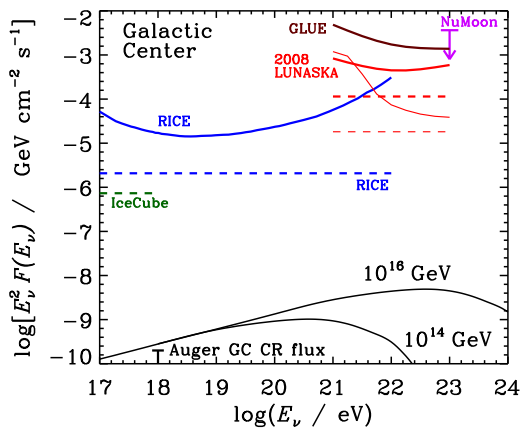}
\caption{Neutrino flux limits for the Galactic Centre from our
2008 LUNASKA ATCA experiments, and based on the GLUE, NuMoon and RICE
experiments, and published IceCube limit. Neutrino limits are
shown as integrated limits to a $E_{\nu}^{-2}$ spectrum (dashed
lines), as well as model-independent limits (solid curves) where
available -- the NuMoon limit is model-independent.  Thin curves
for the LUNASKA ATCA result are for modelling using a toy model of
small-scale lunar roughness. SHDM models (see text) with
$M_X=10^{14}$ GeV and $M_X=10^{16}$ GeV (as labeled).  Limit at
$10^9$~GeV: Auger limit on the {\em cosmic ray} flux from a point
source at the Galactic Centre \citep{Auger0607382}.  }
\label{FluxLimitsGC}
\end{center}
\end{figure}

In Fig.~\ref{FluxLimitsGC} we show all-flavor neutrino flux
limits for the Galactic Center.  We consider the
possibility that the Galaxy's dark matter halo is composed partly
of super-heavy dark matter, which decays or annihilates into particles which cascade
into neutrinos, photons, and nucleons.  We use fragmentation
functions of
\citet{AloisioBerezinskyKachelriess094023} for $M_X=10^{16}$ GeV
and normalize the sum of the gamma-ray plus nucleon components to
the Auger limit on the {\em cosmic ray} flux from a point source
at the Galactic Centre \citep{Auger0607382}.  We plot the
corresponding neutrino flux shown as the curve labeled
``$10^{16}$ GeV''.  We also show the expected neutrino flux
assuming $M_X=10^{14}$ GeV.  Of course whether or not the
neutrino (and cosmic ray) flux from SHDM annihilation appears
point-like will depend on the radial distribution of the dark
matter halo
\citep{EvansFarrarSarkar0103085}.  If this distribution is cusped
the angular distribution will be narrow, approximating to a point
source, but in the case of other models the angular distribution
can be broadened to up to $\sim$$60^\circ$ half width.

In comparison with experiments such as ANITA, our methods to
improve sensitivity to certain specifically targeted regions were
successful.  We have reported the first neutrino flux limits
above $10^{21}$ eV from Centaurus~A and the Galactic Center.  For
Centaurus A, the limit from our 2008 LUNASKA experiments using
the ATCA is currently the strongest at primary neutrino energies
of $5\times 10^{21}$~eV and above.  While not ruling out any of the current
theories of UHE CR production we have proved the viability of the
lunar Cherenkov technique for targeted observation of specific
UHE neutrino source candidates.

\section*{Acknowledgments}

The Australia Telescope Compact Array is part of the Australia
Telescope which is funded by the Commonwealth of Australia for
operation as a National Facility managed by CSIRO. This research
was supported by the Australian Research Council's Discovery
Project funding scheme (project numbers DP0559991 and DP0881006).
J.A-M.\ thanks Xunta de Galicia (INCITE09 206 336 PR) and
Conseller\'\i a de Educaci\'on (Grupos de Referencia Competitivos
-- Consolider Xunta de Galicia 2006/51), and Ministerio de
Ciencia e Innovaci\'on (FPA 2007-65114 and Consolider CPAN) for
financial support, and CESGA (Centro de SuperComputaci\'on de
Galicia) for computing resources and assistance.

\label{lastpage}


\begin{thebibliography}{}

\bibitem[\protect\citeauthoryear{Abbasi et al.}{2009}]{IceCube_CenA} Abbasi R.\ et al., 2009, Phys.\ Rev.\ Lett., 103, 221102

\bibitem[\protect\citeauthoryear{Abraham et al.}{2007a}]{AugerScience07} Abraham J.\  {et al.}  (Auger Collaboration), 2007, Science {318}, 938

\bibitem[\protect\citeauthoryear{Abraham et al.}{2007b}]{Auger0607382}  Abraham J.\  {et al.}\ (Auger Collaboration), 2007, Astropart.\  Phys.\  {27}, 244

\bibitem[\protect\citeauthoryear{Aloisio et al.}{2004}]{AloisioBerezinskyKachelriess094023} Aloisio R. ,  Berezinsky V.\ \& Kachelriess M., 2004, Phys.\  Rev.\  D, {69}, 094023 

\bibitem[\protect\citeauthoryear{\'Alvarez-Mu\~{n}iz}{2006}]{Alvarez-Muniz06}  \'Alvarez-Mu\~{n}iz J., Marques E., Vazquez R.~A., Zas E., 2006, Phys.\  Rev.\  D {74}, 023007


\bibitem[\protect\citeauthoryear{Askar'yan}{1962}]{Askaryan62}  Askar'yan G.~A., 1962, Sov.\  Phys.\  JETP, {14}, 441

\bibitem[\protect\citeauthoryear{Askar'yan}{1965}]{Askaryan65} Askar'yan G.~A., 1965, Sov.\  Phys.\  JETP, {48}, 988


\bibitem[\protect\citeauthoryear{Barwick et al.}{2006}]{ANITA}  Barwick  S.\  W.\ {et al.}, 2006, Phys.\ Rev.\ Lett., {96}, 171101

\bibitem[\protect\citeauthoryear{Benford \& Protheroe}{2008}]{BenfordProtheroe2008MNRAS.383..663B} Benford G., Protheroe R.~J., 2008, MNRAS, 383, 663 

\bibitem[\protect\citeauthoryear{Berezinsky \& Zatsepin}{1977}]{BerezinskyGZK} Berezinsky V.~S. \& Zatsepin G.~T., 1977, Sov.\  J.\  Uspekhi, 20, 361

\bibitem[\protect\citeauthoryear{Besson}{2008}]{Besson08_private}  Besson D., 2008, private communication

\bibitem[\protect\citeauthoryear{Buitink}{2009}]{Buitink09} Buitink S., Radio emission from cosmic particle cascades (Doctoral Thesis), Radboud University Nijmegen, 2009

\bibitem[\protect\citeauthoryear{Cuoco \& Hannestad}{2008}]{CuocoHannestad023007} Cuoco  A.\  \&  Hannestad S., 2008, Phys.\  Rev.\  D, {78}, 023007

\bibitem[\protect\citeauthoryear{Dagkesamanskii \&  Zheleznykh}{1989}]{Dagkesamanskii} Dagkesamanskii R.\  D.,  Zheleznykh I.\ M., 1989, Sov.\  Phys.\  JETP Lett.\  {50}, 233

\bibitem[\protect\citeauthoryear{Ekers et al.}{2009}]{Ekers08}  Ekers R.D.,  James C.W.,  Protheroe R.J.,  McFadden R.A., 2009, Nucl.\ Instr.\ and Meth.\ in Phys.\ Res.\ A, {604}, S106

\bibitem[\protect\citeauthoryear{Engel et al.}{2001}]{Engel01}  Engel R., Seckel D., Stanev T., 2001, Phys.\  Rev.\  D {64}, 093010

\bibitem[\protect\citeauthoryear{Evans et al.}{2002}]{EvansFarrarSarkar0103085} Evans  N.\  W., Ferrer F.,
 Sarkar S., 2002, Astroparticle Physics, {17}, 319

\bibitem[\protect\citeauthoryear{Falcke et al.}{2004}]{Falcke2004_SKAscienceCase}  Falcke H.,  Gorham P.,  Protheroe R.J., 2004, New~Astron.\  Rev.\  {48}, 1487

\bibitem[\protect\citeauthoryear{Feain et al.}{2009}]{Feain_CenA_LobesB_2009ApJ...707..114F} Feain I.~J., et al., 2009, ApJ, 707, 114 

\bibitem[\protect\citeauthoryear{Gorham et al.}{2004}]{GLUE} Gorham  P.\  W.\  {et al.}, 2004, Phys.\  Rev.\  Lett., {93}, 041101
\bibitem[\protect\citeauthoryear{Gorham et al.}{2010}]{ANITA_2010} Gorham P.~W.\ et al., 2010, preprint  arXiv:1003.2961

\bibitem[\protect\citeauthoryear{Greisen}{1966}]{Greisen}  Greisen K., 1966, Phys.\  Rev.\  Lett.\   {16}, 748

\bibitem[\protect\citeauthoryear{Hankins et al.}{1996}]{Parkes}  Hankins T.H.,  Ekers R.D.,  O'Sullivan J.D., 1996, Mon.\  Not.\  Royal Astron.\  Soc.,  {283}, 1027

\bibitem[\protect\citeauthoryear{Hardcastle et al.}{2009}]{Hardcastle_CenA_2009MNRAS.393.1041H} Hardcastle M.~J., Cheung C.~C., Feain I.~J., Stawarz {\L}., 2009, MNRAS, 393, 1041 

\bibitem[\protect\citeauthoryear{James}{2009}]{James_PhD2009} James C.W., Ultra-high energy particle detection with the lunar Cherenkov technique (Doctoral Thesis), University of Adelaide, 2009; available from \verb!http://digital.library.adelaide.edu.au/! \verb!dspace/handle/2440/57706!

\bibitem[\protect\citeauthoryear{James et al.}{2010}]{James_PRD2009} C.\  W.\  James, Ekers R.D., \'Alvarez-Mu\~{n}iz J., Bray J.D., McFadden R.A., Phillips C.J, Protheroe R.J., Roberts P., 2010, Phys.\  Rev. D, vol. 81, Issue 4, id. 042003

\bibitem[\protect\citeauthoryear{James \&  Protheroe}{2009a}]{JamesProtheroe09a}  James C.\  W.,  Protheroe R.\  J., 2009, Astropart.\  Phys., {30}, 318

\bibitem[\protect\citeauthoryear{James \&  Protheroe}{2009b}]{JamesProtheroe09b}  James C.\  W.,  Protheroe R.\  J., 2009, Astropart.\  Phys., 31, 392

\bibitem[\protect\citeauthoryear{Kachelriess et al.}{2009a}]{Kachelriess0805.2608} Kachelriess M., Ostapchenko S., Tomas R., 2009, New J. of Phys., 11, 065017
\bibitem[\protect\citeauthoryear{Kachelriess et al.}{2009b}]{Kachelriess0904.0590} Kachelriess M.,  Ostapchenko  S.,  Tomas R., 2009,  Int.J.Mod.Phys.D18:1591-1595

\bibitem[\protect\citeauthoryear{Kravchenko et al.}{2006}]{RICE} Kravchenko  I.\   {et al.}, 2006, Phys.\  Rev. D {73}, 082002

\bibitem[\protect\citeauthoryear{Mannheim et al.}{2001}]{MannheimProtheroeRachen01} Mannheim K.,  Protheroe R.J.,  Rachen J.P.,2001, Phys.\  Rev.\  D {63}, 023003

\bibitem[\protect\citeauthoryear{M{\"u}cke et al.}{2000}]{SOPHIA2000CoPhC.124..290M} M{\"u}cke A., Engel R., Rachen J.~P., Protheroe R.~J., Stanev T., 2000, CoPhC, 124, 290 

\bibitem[\protect\citeauthoryear{Olhoeft \& Strangway}{1975}]{OlhoeftStrangway75} Olhoeft G.R., Strangway D.W., 1975, Earth and Plan.Sci.Lett., {24}, 394

\bibitem[\protect\citeauthoryear{Protheroe}{2004}]{Protheroe04}  Protheroe R.J., 2004, Astropart.\  Phys.\  {21}, 415

\bibitem[\protect\citeauthoryear{Protheroe \& Clay}{2004}]{ProtheroeClay2004}  Protheroe R.J.,  Clay R.W., 2004, PASA {21}, 1

\bibitem[\protect\citeauthoryear{Protheroe \&  Johnson}{1996}]{ProtheroeJohnson96}  Protheroe R.J.,  Johnson P.A., 1996, Astropart.\  Phys.\  {4}, 253

\bibitem[\protect\citeauthoryear{Rieger \& Aharonian}{2009}]{RiegerAharonian2009A&A...506L..41R} Rieger F.~M., Aharonian F.~A., 2009, A\&A, 506, L41 

\bibitem[\protect\citeauthoryear{Stecker}{1973}]{Stecker1973Ap&SS..20...47S} Stecker F.~W., 1973, Ap\&SS, 20, 47 

\bibitem[\protect\citeauthoryear{Stecker}{1979}]{Stecker1979ApJ...228..919S} Stecker F.~W., 1979, ApJ, 228, 919 

\bibitem[\protect\citeauthoryear{Saltzberg et al.}{2001}]{Saltzberg_GorhamSAND01}  Saltzberg D., Gorham P., Walz D., Field, C., Iverson R., Odian A., Resch G., Schoessow P., Williams D., 2001, Phys.\  Rev.\  Lett.\   {86}, 2802

\bibitem[\protect\citeauthoryear{Scholten et al.}{2009}]{NuMoon} Scholten O., Buitink S., Bacelar J., Braun R., de Bruyn A.G., Falcke H., Singh K., Stappers B., Strom R.G., al Yahyaoui R., 2009, Phys.\ Rev.\ Lett.\ {103}, 191301

\bibitem[\protect\citeauthoryear{Tiffenberg et al.}{2009}]{Auger_CenA_nu} Tiffenberg J. et al. (Auger Collaboration), 2009, Proc. of the 31st Intl. Cosmic Ray Conf., Lodz

\bibitem[\protect\citeauthoryear{Zatsepin \&  Kuzmin}{1966}]{Zatsepin_Kuzmin} Zatsepin G.T.,   Kuzmin V.A., 1966, JETP Lett.\  {4}, 78

\end{thebibliography}
\end{document}